\documentclass[conference]{IEEEtran}

\usepackage{cite}      

\usepackage{graphicx}  

\usepackage{subfigure} 

\usepackage{fancyheadings}
\begin{document}

\title{Smoothed Particle Hydrodynamics in Thermal Phases of a One Dimensional Molecular Cloud}

\author{\authorblockN{Mohsen Nejad-Asghar}
\authorblockA{Department of Physics,\\
Damghan University of Basic Sciences,\\
Damghan, Iran\\
nasghar@dubs.ac.ir} \and
\authorblockN{Diego Molteni}
\authorblockA{Dipartimento di Fisica e Tecnologie Relative,\\
Universita di Palermo, Viale delle Scienze, 90128,\\
Palermo, Italy\\
molteni@unipa.it} }


\maketitle

\begin{abstract}

We present an investigation on effect of the ion-neutral (or
ambipolar) diffusion heating rate on thermal phases of a
molecular cloud. We use the modeling of ambipolar diffusion with
two-fluid smoothed particle hydrodynamics, as discussed by
Nejad-Asghar \& Molteni. We take into account the ambipolar drift
heating rate on the net cooling function of the molecular clouds,
and we investigate the thermal phases in a self-gravitating
magnetized one dimensional slab. The results show that the
isobaric thermal instability criterion is satisfied in the outer
parts of the cloud, thus, these regions are thermally unstable
while the inner part is stable. This feature may be responsible
for the planet formation in the outer parts of a collapsing
molecular cloud and/or may also be relevant for the formation of
star forming dense cores in the clumps.

\end{abstract}

\section{Introduction}

In most plasmas, the forces acting on the ions are different from
those acting on the electrons, so naively one would expect one
species to be transported faster than the other, by diffusion or
convection or some other process. If such differential transport
has a divergence, then it will result in a change of the charge
density, which will in turn create an electric field that will
alter the transport of one or both species in such a way that
they become equal. In astrophysics, \textit{ambipolar diffusion}
refers specifically to the decoupling of neutral particles from
plasma. The neutral particles in this case are mostly hydrogen
molecules in a cloud, and the plasma is composed of ions (mostly
protons) and electrons, which are tied to the interstellar
magnetic field.

Nejad-Asghar (2007) has recently made the assumption that the
molecular cloud is initially an uniform ensemble which then
fragments due to thermal instability. He finds that ambipolar
drift heating is inversely proportional to density and its value,
in outer parts of the cloud, can be significantly larger than the
average heating rates of cosmic rays and turbulent motions. His
results show that the isobaric thermal instability can occur in
outer regions of the cloud.  The rapid growth of thermal
instability results a strong density imbalance between the cloud
and the low-density surroundings; therefore it may produce the
cloud fragmentation and formation of the condensations.

Many authors have developed computer codes that attempt to model
ambipolar diffusion. Black \& Scott~(1982) used a two-dimensional,
deformable-grid algorithm to follow the collapse of isothermal,
non-rotating magnetized cloud. The three-dimensional work of
MacLow et al. (1995) treats the two-fluid model in a version of
the \textsc{zeus} magnetohydrodynamic code. An algorithm capable
of using the smoothed particle hydrodynamics (SPH) to implement
the ambipolar diffusion in a fully three-dimensional,
self-gravitating system was developed by Hosking \& Whitworth
(2004). They described the SPH implementation of two-fluid
technique that was tested by modeling the evolution of a dense
core, which is initially thermally supercritical but magnetically
subcritical. Nejad-Asghar \& Molteni (2007) have recently
optimized the two-fluid SPH implementation to test the pioneer
works on the behavior of the ambipolar diffusion in an isothermal
self-gravitating molecular layer.

In this paper, we include the SPH equivalent of the energy
equation, which follows from the first law of thermodynamics.
Here, we include the ambipolar drift heating rate in the net
cooling function, and we investigate the thermal phases in the
self-gravitating magnetized molecular layer. For this purpose,
the two-fluid SPH technique and its algorithm are given in
section~2. Section~3 devotes to the basic equations of the one
dimensional molecular layer and its evolution by SPH. The summary
and conclusion are presented in section~4.

\section{Numerical method}

In two-fluid SPH technique of Hosking \& Whitworth (2004), the
initial SPH particles represent by two sets of \textit{molecular
particles}: magnetized ion SPH particles and non-magnetized
neutral SPH particles. In this method, for each SPH particle we
must create two separate neighbor lists: one for neighbors of the
same species and another for those of different species.
Consequently, each particle has two different smoothing lengths.
In the following sections we refer to neutral particles as
$\alpha$ and $\beta$, and ion particles as $a$ and $b$; the
subscripts $1$ and $2$ refer to both ions and neutral particles.
We adopt the usual smoothing spline-based kernel (Monaghan \&
Lattanzio 1985) and apply the symmetrized form proposed by
Hernquist \& Katz (1989)
\begin{equation}
W_{12}= W(\mid\textbf{r}_1 -\textbf{r}_2\mid, h_{12})
\end{equation}
where $\textbf{r}_1$, $\textbf{r}_2$ are positions of the
particles $1$ and $2$, respectively. $h_{12}$ is the smoothing
length of particle $1$ when considering neighbors of kind $2$. We
update the adaptive smoothing length according to the octal-tree
based nearest neighbor search algorithm (NNS).

The exact fraction of the total fluid that is ionized depends upon
many factors (e.g. the neutral density, the cosmic ray ionization
rate, how efficiently ionized metals are depleted on to dust
grains). Here, we use the expression employed by Fiedler \&
Mouschovias (1992), which states that for $10^{8} < n < 10^{15}
\mathrm{m}^{-3}$,
\begin{equation}\label{ionden}
  \rho_i=\epsilon (\rho^{1/2} + \epsilon' \rho^{-2}),
\end{equation}
where in standard ionized equilibrium state, $\epsilon\sim
7.5\times 10^{-15} \mathrm{kg^{1/2}.m^{-3/2}}$ and $\epsilon'
\sim 4\times 10^{-44} \mathrm{kg^{5/2}.m^{-15/2}}$ are valid. In
reality, the gas in this case is very weakly ionized, thus, we
adopt the approximation $\rho=\rho_n+\rho_i\approx\rho_n$ in
fluid equations. In this case, the molecular cloud is considered
as global neutral which consists of a mixture of atomic and
molecular hydrogen (with mass fraction $X$), helium (with mass
fraction $Y$), and traces of $\mathrm{CO}$ and other rare
molecules, thus, the mean molecular weight is given by
$1/\mu=X/2+Y/4$.

The neutral density in place of neutral particles is estimated via
usual summation over neighboring neutral particles
\begin{equation}
\rho_{n,\alpha}=\sum_\beta m_\beta W_{\alpha\beta},
\end{equation}
while in place of ions, $\rho_{n,a}$, is given by interpolation
technique from the values of the nearest neighbors. The ion
density is evaluated via equation (\ref{ionden}) as follows
\begin{equation}
\rho_{i,1}=1.8 \times 10^{-9} \rho_{n,1}^{1/2} (1+3.5 \times
10^{-17} \rho_{n,1}^{-5/2}),
\end{equation}
for both places of ions and neutral particles. According to this
new ion density, we update the mass of ion $a$ as
\begin{equation}
m_a^{new}=m_a^{old} \frac{\rho_a^{new}}{\rho_a^{old}},
\end{equation}
in each time step so that the usual summation law
\begin{equation}\label{ionsum}
\rho_{i,a}=\sum_\beta m_b W_{ab},
\end{equation}
for ions might being appropriate.

The SPH form of the drift velocity of ion particle $a$ is given
by Hosking \& Whitworth (2004) as
\begin{eqnarray}\label{driftsph}
\nonumber v_{d,a}= \frac{1}{\gamma_{AD}\rho_{n,a}}[-\frac{1}{\mu_0
\rho_{i,a}} \sum_b \frac{m_b}{\rho_{i,b}} B_b B_a
\frac{dW_{ab}}{dz_a} ~~~~~~~~\\ - \sum_b m_b \Pi_{ab}
\frac{dW_{ab}}{dz_a}].
\end{eqnarray}
where $\Pi_{ab}$ is the artificial viscosity between ion
particles $a$ and $b$
\begin{equation}
\Pi_{ab}=\cases{
       \frac{\alpha^* a_{ab} \mu_{ab}
       +\beta^* \mu_{ab}^2}{\bar{\rho}_{ab}}, &
       if $\textbf{v}_{ab}.\textbf{r}_{ab}<0$,\cr
       0 , & otherwise,}
\end{equation}
where $\bar{\rho}_{ab}= \frac{1}{2}(\rho_a+\rho_b)$ is an average
density, $\alpha^*\sim 1$ and $\beta^*\sim 2$ are the
coefficients, $a_{ab}= \frac{1}{2} (a_a+a_b)$ is the mean sound
speed, and $\mu_{ab}$ is defined as its usual form
\begin{equation}
\mu_{ab}=-\frac{\textbf{v}_{ab}
\cdot\textbf{r}_{ab}}{\bar{h}_{ab}}
\frac{1}{r_{ab}^2/\bar{h}_{ab}^2+0.001}
\end{equation}
with $\bar{h}_{ab}= \frac{1}{2}(h_a+h_b)$. Nejad-Asghar, Khesali
\& Soltani (2008) has recently considered the coefficients in the
Monaghan's standard artificial viscosity as time variable, and a
restriction on them is proposed such that avoiding the undesired
effects in the subsonic regions. Here, we use the Monaghan's
standard artificial viscosity, since the cloud contraction is
quasi-hydrostatic and there is not supersonic motions and shock
formation during this contraction.

It is better to keep in mind the second golden rule of SPH which
is to rewrite formulae with the density inside operators
(Monaghan~1992). In this case, the drift velocity is interpolated
as
\begin{eqnarray}\label{driftsph2}
\nonumber v_{d,a}=
\frac{1}{\gamma_{AD}\rho_{n,a}}[-\frac{1}{2\mu_0 \rho_{i,a}}
\sum_b \frac{m_b}{\rho_{i,b}} (B_b^2-B_a^2) \frac{dW_{ab}}{dz_a}
~~\\- \rho_{i,a} \sum_b \frac{m_b}{\rho_{i,b}} \Pi_{ab}
\frac{dW_{ab}}{dz_a}],
\end{eqnarray}
where two extra density terms are introduced, one outside and one
inside the summation sign. This comes as a result of the
approximation to the volume integral needed to perform function
interpolation (Nejad-Asghar \& Molteni~2007).

There is no analytical expression that allows to calculate the
value of drift velocity of the neutral particles. Here, we use the
interpolation technique that starts at the nearest neighbor, then
add a sequence of decreasing corrections, as information from
other neighbors is incorporated (e.g., Press et al. 1992). These
drift velocities at neutral places are used to estimate the drag
acceleration
\begin{equation}\label{dragaccsph}
  a_{drag,\alpha} = \gamma_{AD} \rho_{i,\alpha} v_{d,\alpha},
\end{equation}
instead the method of Hosking \& Whitworth (2004) who used the
expression of Monaghan \& Kocharayan~(1995). In the usual
symmetric form, the self-gravitating SPH acceleration equation
for neutral particle $\alpha$ is
\begin{equation}\label{accneut}
\frac{dv_\alpha}{dt}=g_\alpha-\sum_\beta m_\beta
(\frac{p_\alpha}{\rho^2_\alpha} +\frac{p_\beta}{\rho^2_\beta} +
\Pi_{\alpha\beta} )\frac{d W_{\alpha\beta}}{d z_\alpha}
+a_{drag,\alpha}
\end{equation}
where $g_\alpha$ is the gravitational acceleration of particle
$\alpha$.

The ion momentum equation assuming instantaneous velocity update
so that we have
\begin{equation}
v_a= \sum_\beta \frac{m_\beta}{\rho_\beta} v_\beta W_{a\beta}
+v_{d,a}
\end{equation}
where the first term on the right-hand side gives the neutral
velocity field at the ion particle $a$, calculated using a
standard SPH approximation. Finally, the magnetic induction
equation in SPH is
\begin{equation}
\frac{dB_a}{dt}= \sum_b \frac{m_b}{\rho_b} B_a v_{ab} \frac{d
W_{ab}}{dz_a}.
\end{equation}
where the usual notation is used.

\begin{figure}
\centering
\includegraphics[width=3.2in]{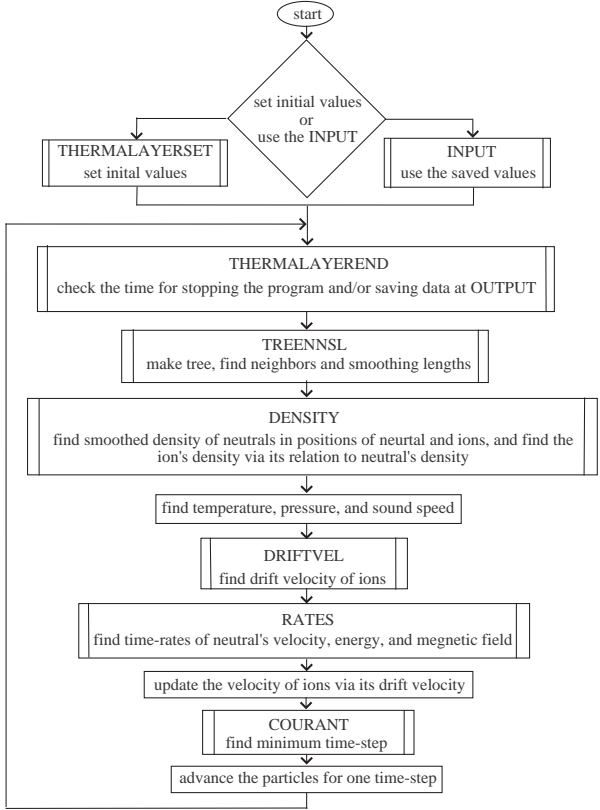}
\caption{Algorithm of the two-fluid smoothed particle
hydrodynamics for simulation of thermal phases in a one
dimensional molecular cloud.} \label{algor}
\end{figure}

In ambipolar diffusion process, the ion particles are physically
diffused through the neutral fluid, thus, the ions will be bared
in the boundary regions of the cloud (i.e. without any neutral
particles in their neighbors). We check the position of ions
before making a tree and nearest neighbor search, so that we send
out the bared ion particles in the boundary regions of the
simulation at next time-step. Both the cloud and boundary regions
contain ion and neutral particles. We set up boundary particles
($4h_1$ up and down in $z$) using the linear extrapolation
approach (from the values of the inner particles) to attribute
the appropriate drift velocity, drag acceleration, pressure
acceleration, and the magnetic induction rate to the boundary
particles. The algorithm of the two-fluid SPH for simulation of
thermal phases in a one dimensional molecular cloud is shown in
Fig.~\ref{algor}.

\section{layer evolution}

\subsection{basic equations}

We write the continuity equation of neutral part as its common
form
\begin{equation}\label{masscon}
\frac{d \rho}{d t} =- \rho \frac{\partial v}{\partial z},
\end{equation}
while, the relation (\ref{ionden}) is used to determine the ion
density wherever it is required in the two-fluid equations. The
momentum equation of neutral part then becomes
\begin{equation}\label{momcon}
  \frac{d v}{d t} = g - \frac{1}{\rho} \frac{\partial}{\partial z} (p +
  \frac{B^2}{2\mu_0})
\end{equation}
where the gravitational acceleration $g$ obeys the poisson's
equation
\begin{equation}\label{poisson}
  \frac{\partial g}{\partial z} = -4 \pi G \rho,
\end{equation}
and the pressure is given by the ideal gas equation of state
\begin{equation}\label{pressu}
  p=\frac{k_B}{\mu m_\mathrm{H}}\rho T
\end{equation}
where $m_\mathrm{H}$ is the mass of hydrogen atom and the
temperature $T$ is approximated the same as for both neutral and
ion fluids ($T_i=T_n=T$).

The thermal energy per unit mass is
\begin{equation}\label{ut}
u(T)=(\frac{5}{4}X + \frac{3}{8}Y) \frac{k_B T}{ m_H}
\end{equation}
where the mean internal (rotation and vibration) energy of an
$\mathrm{H_2}$ molecule is included. The energy equation follows
from the first law of thermodynamics, that is
\begin{equation}\label{energycon}
  \frac{du}{dt} = - \frac{p}{\rho} \frac{\partial v}{\partial z} - \Omega_{(\rho,T)},
\end{equation}
where $\Omega_{(\rho,T)}$ is the net cooling function
\begin{equation}\label{netcool}
  \Omega_{(\rho,T)} \equiv \Lambda_{(n)} (\frac{T}{10 \mathrm{K}})^{\beta_{(n)}} -
  (\Gamma_{CR} + \Gamma_{AD}),
\end{equation}
where $\Gamma_{CR}$ and $\Gamma_{AD}$ are the heating rates due
to cosmic rays and ambipolar diffusion, respectively, and
$\Lambda_{(n)}$ and $\beta_{(n)}$  are the parameters for the gas
cooling function that here we use the polynomial fitting
functions, outlined by Nejad-Asghar~(2007), as follows
\begin{eqnarray}\label{lambda0}
  \nonumber \log\left(\frac{\Lambda_{(n)}}{\mathrm{J.kg^{-1}.s^{-1}}}\right) = -8.98 -
  0.87 (\log \frac{n}{n_0}) \\ - 0.14 (\log \frac{n}{n_0})^2,
\end{eqnarray}
\begin{equation}\label{beta}
  \beta_{(n)} = 3.07 - 0.11 (\log \frac{n}{n_0}) - 0.13 (\log \frac{n}{n_0})^2,
\end{equation}
where $n_0=10^{12} \mathrm{m^{-3}}$.

The magnetic fields are directly evolved by charged fluid
component, as follows:
\begin{equation}\label{magcon}
  \frac{d B}{d t} = - B \frac{\partial v}{\partial z} +
  \frac{\partial}{\partial z} (B v_d),
\end{equation}
where the last term outlines the ambipolar diffusion effect with
drift velocity,
\begin{equation}\label{drift}
  v_d = -\frac{1}{\gamma_{AD} \epsilon \rho \rho_i} \frac{\partial}{\partial
  z} (\frac{B^2}{2\mu_0}),
\end{equation}
which is obtained by assumption that the pressure and
gravitational force on the charged fluid component are negligible
compared to the Lorentz force because of the low ionization
fraction. The notation $\gamma_{AD} \sim 3.5 \times 10^{10}
\mathrm{m^3/kg.s}$ in drift velocity represents the collision drag
coefficient.

\subsection{results}

The chosen physical scales for length and time are $[l]=200
\mathrm{AU}$, and $[t]=10^3 \mathrm{yr}$, respectively, so that
velocity unit is approximately $[v]=1 \mathrm{km.s^{-1}}$. The
Newtonian constant of gravitation is set $G= 1 [m]^{-1} [l]^3
[t]^{-2}$ for which the calculated mass unit is $[m]= 4.5 \times
10^{29} \mathrm{kg}$. Consequently, the derived physical scale for
density, energy per unit mass, and drag coefficient are $[\rho]=
1.7\times 10^{-11} \mathrm{kg.m^{-3}}$, $[u]= 10^{6}
\mathrm{J.kg^{-1}}$, and $\gamma_{AD}=1.8\times 10^{10} [l]^3
[m]^{-1} [t]^{-1}$, respectively. In this manner, the numerical
values of $\epsilon$ and $\epsilon'$ are $1.8 \times 10^{-9}
[l]^{-3/2} [m]^{1/2}$ and $3.5 \times 10^{-17} [l]^{-15/2}
[m]^{5/2}$, respectively. The magnetic field is scaled in units
such that the constant $\mu_0$ is unity. Since the magnetic flux
density has dimensions
\begin{equation}\label{magdimen}
  [B]=\frac{[m]}{[t][charge]},
\end{equation}
while $\mu_0$ has dimensions
\begin{equation}\label{mudimen}
  [\mu_0]=\frac{[m][l]}{[charge]^2},
\end{equation}
specifying $\mu_0=1$ therefore scales the magnetic field equal to
$[B]=5.1 \mathrm{nT}$. With aforementioned units, the thermal
energy per unit mass (\ref{ut}) is represented by
$8.3\times10^{-3} (5X/4+3Y/8) T$, the heating rates due to cosmic
rays and ambipolar diffusion are $\Gamma_{CR} = 1.1 \times
10^{-3} [u]/[t]$ and
\begin{equation}\label{adheat}
  \Gamma_{AD,\alpha}= \gamma_{AD}\rho_{i,\alpha} v_{d,\alpha}^2,
\end{equation}
respectively, and the parameters for the gas cooling function are
\begin{eqnarray}\label{lambda02}
  \nonumber \log\left(\frac{\Lambda_\alpha}{[u]/[t]}\right) = -4.48 -
  0.87 (\log \frac{\rho_\alpha}{2.24\times10^{-4}}) \\ - 0.14 (\log \frac{\rho_\alpha}{2.24\times10^{-4}})^2,
\end{eqnarray}
\begin{eqnarray}\label{beta2}
 \nonumber \beta_\alpha = 3.07 - 0.11 (\log \frac{\rho_\alpha}{2.24\times10^{-4}})
  \\ - 0.13 (\log \frac{\rho_\alpha}{2.24\times10^{-4}})^2.
\end{eqnarray}

\begin{figure*}
\centering
\includegraphics[width=6in]{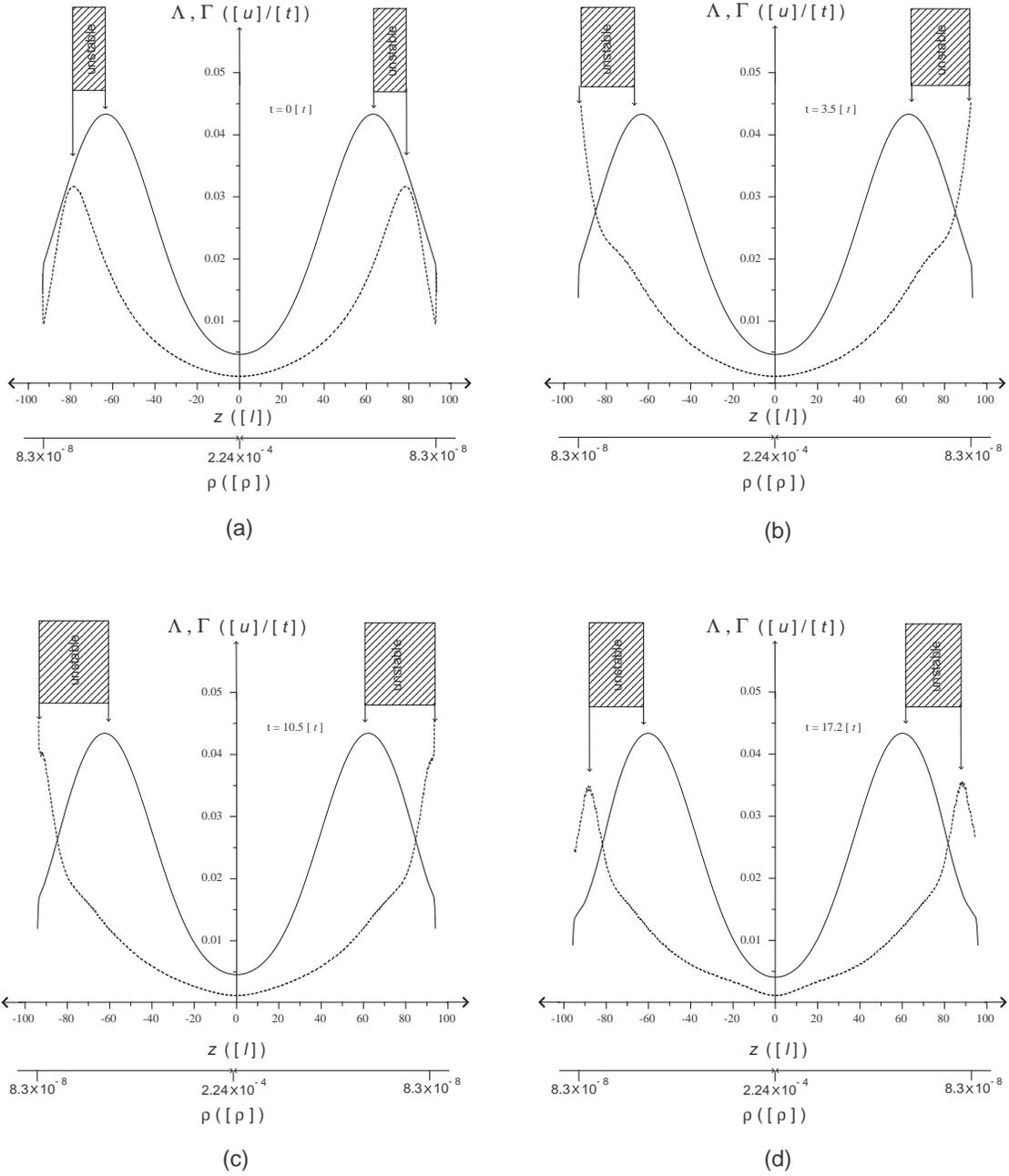}
\caption{The values of the cooling (solid) and heating (dash)
functions versus position (and neutral density), at initial time
and $t=3.5 [t]$, $10.5 [t]$ and $17.2 [t]$, respectively. Thermal
unstable regions are shaded in this figure.} \label{coolheat}
\end{figure*}

The initial conditions for this simulation are a parallel magnetic
field directed perpendicular to the $z$-axis so that the initial
ratio of magnetic to gas pressure is everywhere a constant
($\alpha_0=1$), and a density profile given by
\begin{equation}\label{inidenprof}
    \rho = \frac{\rho_0}{\cosh ^2 (z/z_\infty)}
\end{equation}
where $\rho_0$ is the initial central density and $z_\infty \equiv
c_s \sqrt{ (1+ \alpha_0) / 2 \pi G \rho_0}$ is a length-scale
parameter according to the initial sound speed $c_s$ (see Fig.~1
of Nejad-Asghar \& Molteni~2007). The magnetic field is assumed
to be frozen in the fluid of charged particles and the central
density is assumed to be $\rho_0= 2.24\times10^{-4}[\rho]$. We
choose a molecular cloud with the mass fraction of molecular
hydrogen and helium are $X=0.75$ and $Y=0.25$, respectively, and
have initial uniform temperature of $T_0 =50\mathrm{K}$. We
assume that the cloud layer is spread from $z=-93.2 [l]$ to
$z=+93.2 [l]$ (according to Fig.~1 of Nejad-Asghar \& Molteni
2007). The initial values of the cooling and heating functions
are shown in Figure~\ref{coolheat}a. As presented in this figure,
the isobaric thermal instability criterion,
\begin{equation}\label{thermcri}
  \frac{\partial \Lambda}{\partial \rho} > \frac{\partial \Gamma}{\partial
  \rho},
\end{equation}
is satisfied in the outer parts of the cloud, thus, these regions
are thermally unstable while the inner part is stable as outlined by
Nejad-Asghar~(2007).

The present SPH code has the main features of the TreeSPH class
so that the nearest neighbors searching are calculated by means
of this procedure. We integrate the SPH equations using the one
order simple integration scheme. The selection of time-step,
$\Delta t$, is of great importance. There are several time-scales
that can be defined locally in the system. For each particle $1$,
we calculate the smallest of these time-scales using its smallest
smoothing length, $h_1$, i.e.
\begin{equation}\label{timestep}
\Delta t_1=C_{cour} \min[ \frac{h_1}{| v_1 |} ,
\frac{h_1}{v_{A,1}} , \frac{h_1}{a_1}],
\end{equation}
where $v_A = B / \sqrt{\mu_0 \rho_i}$ is the Alfv\'{e}n speed of
ion fluid and $C_{cour}$ is the Courant number which in this
paper is adopted equal to $0.3$ (for numerical stability). The
evolution were carried out to time $<u> / <\Lambda,\Gamma>\sim
0.43/0.025= 17.2[t]$ so that the stability of the outer parts of
the cloud may be revealed. The values of the cooling and heating
rates versus position (and neutral density) are shown in
Fig.~\ref{coolheat}b-d, at times $t=3.5 [t]$, $10.5 [t]$ and
$17.2 [t]$, respectively.

\section{Summary and conclusions}

In the preceding work of Nejad-Asghar \& Molteni (2007), two-fluid
SPH implementation was used to check the pioneer works on behavior
of the ambipolar diffusion in an isothermal self-gravitating
layer. In this paper, we include the SPH equivalent of the energy
equation that can be obtained from the first law of
thermodynamics. Here, we incorporate the ambipolar drift heating
in the net cooling function, and we investigate the thermal
phases in the self-gravitating magnetized molecular layer.

The values of the cooling and heating functions are shown in
Figure~\ref{coolheat}. According to this figure, the isobaric
thermal instability criterion is satisfied in the outer parts of
the cloud, thus, these regions are thermally unstable while the
inner part is stable. The SPH equations were integrated, and the
evolution were carried out to time $17.2[t]$ so that the
stability of the outer parts of the cloud is revealed.

It is obvious that the instability of the cloud at the outer
parts, causes the formation two relative cool regions in that
area. The rapid growth of thermal instability results a strong
density imbalance between the cloud and the low-density
surroundings. This feature may be responsible for the planet
formation in the outer parts of a collapsing molecular cloud
and/or may also be conscientious for the formation of star
forming dense cores in the clumps.

\section*{Acknowledgment}
The MNA would like to thank the Research Council of the Damghan
University of Basic Sciences.



%

\end{document}